\documentstyle[11pt,newpasp,twoside,epsf]{article}
\reversemarginpar

\begin{document}
\title{Mapping the Dark Energy Equation of State} 
\author{Eric V.\ Linder} 
\affil{Berkeley Lab, 1 Cyclotron Road, M/S 50R5008, 
Berkeley, CA 94720-8158 USA; evlinder@lbl.gov} 

\begin{abstract}
The acceleration of the expansion of the universe 
has deep implications for structure formation, the 
composition of the universe, and its fate.  Roughly 
70\% of the energy density is in a dark energy, 
whose nature remains unknown.  Mapping the 
expansion history through supernovae, mapping the 
geometry of the universe and formation of structure 
through redshift surveys, and mapping the distance 
to recombination through the cosmic microwave 
background provide complementary, precise probes of 
the equation of state of the dark energy.  Together 
these next generation maps of the cosmos can reveal 
not only the value today, but the redshift variation, 
of the equation of state, providing a critical 
clue to the underlying physics. 
\end{abstract}

\section{Introduction}

Observations of the distance-redshift relation of Type Ia 
supernovae have given firm evidence of an accelerated expansion 
of the universe (Knop et al.\ 2003; Tonry et al.\ 2003).  
As calibratable ``standardized 
candles'', these supernovae are excellently suited to map the 
expansion history $a(t)$ due to the direct relation 
between the measurements and the cosmological dynamics.  The 
redshift of the supernova, $z=a^{-1}-1$, measures the scale factor 
$a$, the size of the universe when the supernova exploded relative 
to its current size.  The calibrated peak magnitude gives the 
distance, which translates to the lookback time to the explosion. 

The data clearly indicate an acceleration to the expansion rather 
than the slowing down under gravitational attraction that was 
previously expected.  This gravitational repulsion is generally 
interpreted in terms of an additional component to the energy density 
of the universe, and given the name dark energy.  Since the effective 
gravitating mass in general relativity depends on both the energy 
density $\rho$ and pressure $p$ in the combination $\rho+3p$, such 
a repulsion and hence acceleration could be induced by a component 
with strongly negative pressure.  Characterized in terms of the 
equation of state ratio $w=p/\rho$, the condition for a single 
component to accelerate the expansion is $w<-1/3$. 

In order to achieve the acceleration deduced from the distance-redshift 
measurements, in a flat universe with (decelerating) matter density as well, 
the energy density in dark energy must amount to $\sim$70\% of the total. 
Thus the majority of the universe is composed of dark energy, determining 
the cosmic dynamics and the fate of the universe.  Moreover, the equation 
of state must be substantially negative, $w\approx -1$.  The physics 
underlying the dark energy sets the equation of state, so to understand 
this new gravitational or high energy physics requires precise and 
accurate measurement of this quantity.  Is the dark energy 
Einstein's cosmological constant ($w=-1$ exactly), some high energy 
physics scalar field (often called quintessence), or a sign of modifications 
to gravity or the presence of extra dimensions?  Except for the cosmological 
constant, almost all theories predict dynamical dark energy, with an 
equation of state evolving with the cosmic expansion.  In fact, this 
dynamics, in the form of the time variation $w(z)$, contains the main clue 
to the new physics.  Thus the goal is to bring together astrophysics and 
particle physics to map the dark energy equation of state. 

\section{Mapping the Expansion History} 

The dark energy affects both the expansion history of the universe 
and the growth history of large scale structure.  In addition to the 
supernova measurements of distances to redshifts of order one, the 
location of the acoustic peaks in the cosmic microwave background 
power spectrum provides the distance to the last scattering surface 
at $z=1089$.  Observations of galaxy clustering, the mass power 
spectrum, and velocity distortions of large scale structure depend 
on the growth history.  

At present, no one method provides tight 
constraints on the equation of state (EOS) due to degeneracies between 
cosmological parameters, e.g.\ the dark energy density $\Omega_w$ and 
the equation of state $w$.  But combining cosmological probes can break 
these degeneracies and improve the estimation.  One must be careful 
to ensure that 
the dark energy has been consistently taken into account when using 
the quantities derived from different methods, and that, for example, 
a quoted determination of matter density did not assume a 
cosmological constant universe.  In the current state of the art, 
such a consistent analysis leads to a measure of the  
{\it assumed constant\/} equation of state $w=-1.05^{+0.15}_{-0.20}$ 
(Knop et al.\ 2003).  Note that many models other than the cosmological 
constant possess an averaged EOS near $-1$ for part of their evolution, 
so these limits do not rule out many physically distinct models.  

But the observational situation is rapidly improving.  The approximation 
of a constant $w$ will soon be confronted with large data sets of 
supernovae to $z>1$ and deep galaxy redshift surveys, and of course 
the CMB data probe a quantity $\langle w\rangle$ different from a 
low redshift, averaged $w$.  Furthermore, the time variation $w'$ is 
a critical clue to the underlying fundamental physics.  Analysis of 
these data in terms of an {\it a priori\/} fixed EOS is insufficient 
and can both blind and mislead us. 

Fortunately there exists a simple parametrization of EOS that 
incorporates the dynamical aspects but does not require model 
dependent elements that interfere with comparison of predictions 
among models.  The parametrization 
\begin{equation} 
w(z)=w_0+w_a(1-a)=w_0+w_a\,z/(1+z) 
\end{equation} 
is well behaved to high redshift and serves as an excellent 
approximation (see Fig.\ 1, left panel) to slow roll scalar field 
models of dark energy (Linder 2003a). 

\begin{figure} 
\plottwo{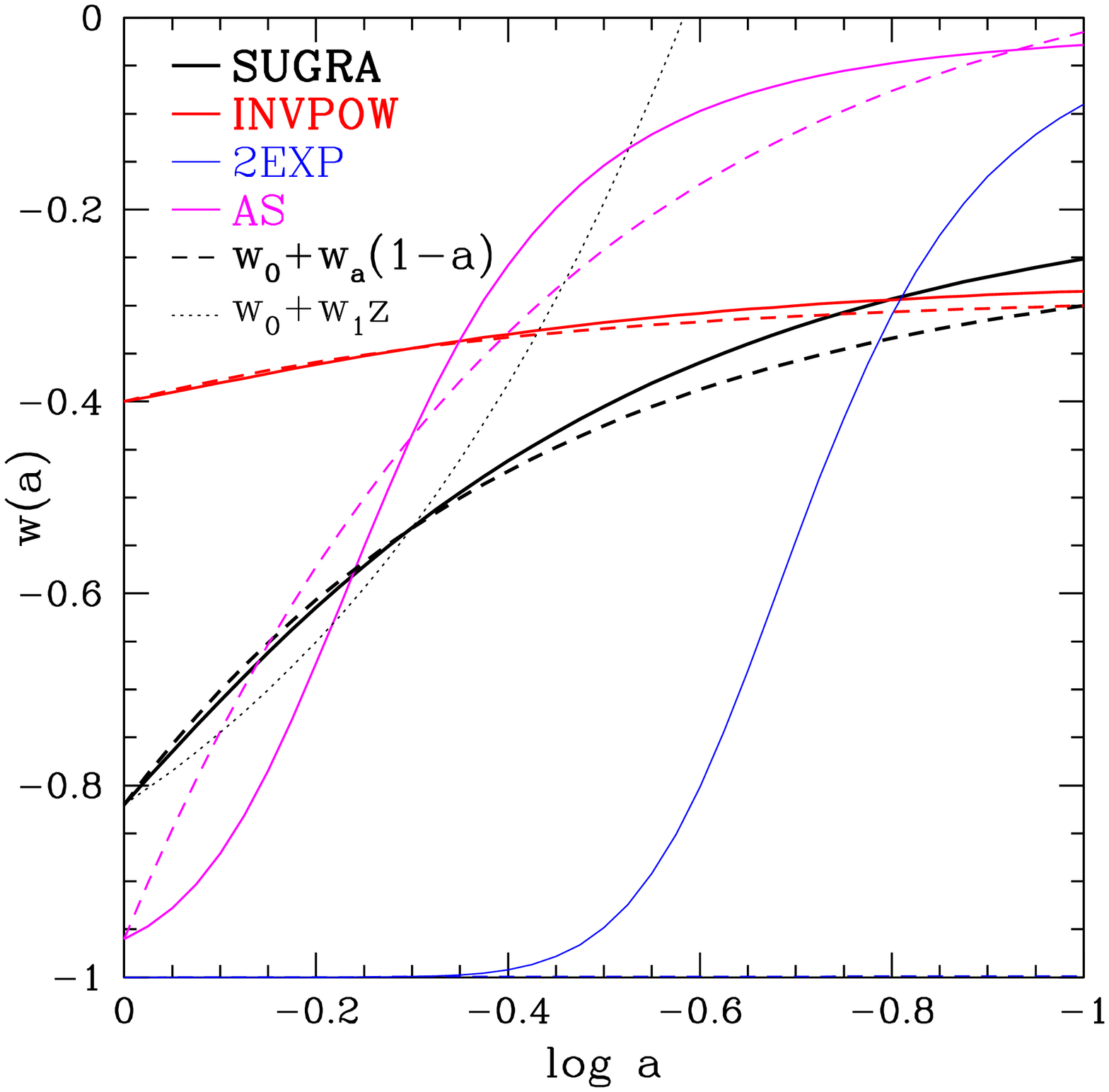}{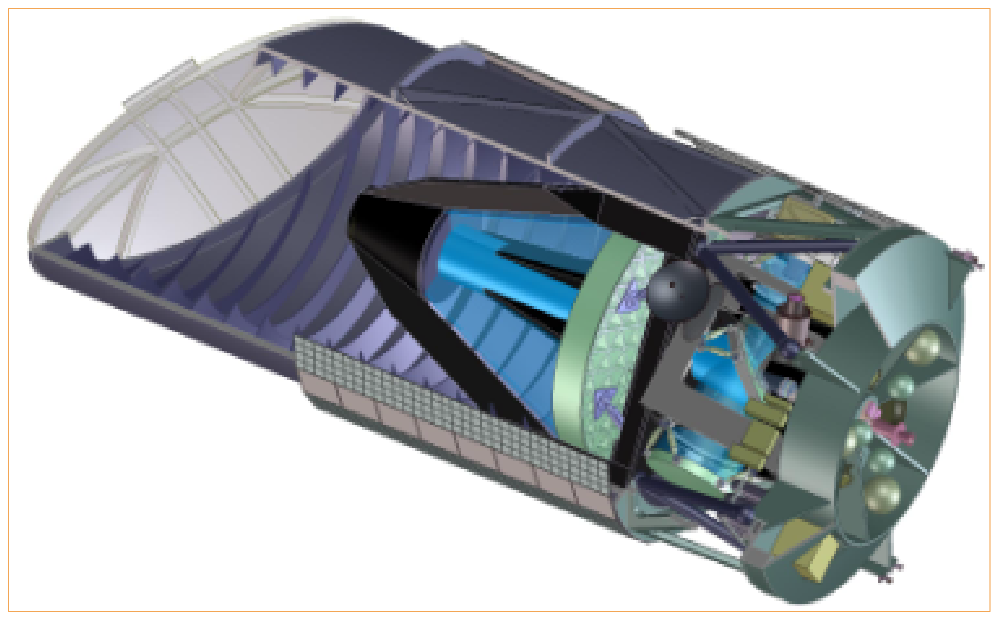}
\caption{{\it Left panel\/} -- The equations of state of four dark 
energy models are plotted as a function of expansion factor.  Dashed 
lines show the reconstruction from the simple parametrization in 
eq.\ 1.  The dotted line gives the old, linear in redshift, parametrization 
for the SUGRA case. {\it Right panel\/} -- Cutaway view of proposed SNAP 
satellite, designed to measure the equation of state and its variation. 
}
\label{fig.wa}
\end{figure}

Moreover, modifications of the Friedmann equation for the rate 
of expansion can be written in terms of an effective $w(z)$.  If 
we admit ignorance of the physical mechanism leading to the observed 
acceleration, then we would write 
\begin{equation} 
[H(z)/H_0]^2=\Omega_M(1+z)^3+\delta H^2/H_0^2, 
\end{equation} 
where we know there exists some matter density $\Omega_M$ and 
allow some additional term $\delta H^2$, which may or may not be a 
real dark energy.  But we can still consistently define an 
effective EOS as 
\begin{equation} 
w(z)\equiv -1 +\frac{1}{3}{d\ln \delta H^2\over d\ln (1+z)}. \label{eq.fried} 
\end{equation} 

So how do we design next generation cosmological probes to uncover 
the crucial information of $w(z)$?  The clearest hope resides in 
observations that involve simple, well understood physics, with 
tightly constrained systematic uncertainties.  Perhaps the most 
promising is the technique that first discovered the dark energy -- 
the Type Ia supernova method.  Each supernova provides not 
just a single data point but a rich stream of crosschecking 
information in the form of its light curve (magnitude vs.\ time) 
and energy spectrum. 

The Supernova/Acceleration Probe (SNAP; see right panel of Fig.\ 1) 
is a proposed mission 
dedicated to studying dark energy, employing the supernova method 
along with other techniques.  It consists of a 2-meter aperture telescope 
in space coupled to a 1 degree field of view mosaic camera 
instrumented with over half a billion pixels, plus a low resolution 
spectrograph.  Nine filters cover the optical and near infrared 
from 3500-17000$\AA$.  SNAP can discover and follow up over 2000 
supernovae in the range $z=0.1-1.7$, characterizing them precisely 
in terms of their spectra, and bounding systematic uncertainties 
below 0.02 mag (1\% in distance). 

In its deep survey mode, SNAP repeatedly scans 15 square degrees of sky 
to study supernovae.  At the same time these observations can be used 
to build up a deep weak gravitational lensing map of the sky, detailing 
the dark matter distribution.  Moreover, the data resources cover 9000 
times the area of a Hubble Deep Field and reach coadded depth of 
AB magnitude 30.3.  The wide field survey images 300 square degrees or 
more to AB=28.1 in nine filters, with the weak lensing information 
providing important constraints on the cosmological parameters 
complementary to the supernova determinations (Refregier et al.\ 2003). 

Indeed, complementarity between precision methods greatly strengthens 
the confidence in and leverage of the dark energy parameter estimations. 
For such a momentous discovery as dark energy, we need to place a 
premium on accurate observations, where systematics can be well understood 
and tightly limited.  But even so, the use of two or more 
techniques with distinct sources of systematics should be strongly 
sought to ensure dependable conclusions. 

Figure 2 (left panel) shows the advantages to combining supernova 
data from SNAP with CMB data from Planck.  While SNAP alone does 
constrain dark energy models, the inclusion of CMB data means that no 
prior knowledge on the matter density is required.  This makes the 
conclusions much cleaner, and we see that the parameter contours are 
much tighter as well, equivalent to those with a prior $\sigma(\Omega_M) 
\approx0.01$ (Frieman et al.\ 2003).  Together the two data sets can 
detect the time variation $w'$ from, say, a supergravity inspired 
dark energy model at the 99\% confidence 
level.  This means we will have advanced from originally detecting the 
mere existence of dark energy (that $\Omega_\Lambda>0$) at 99\% probability, 
to characterizing its EOS dynamics at the same level. 

\begin{figure} 
\plottwo{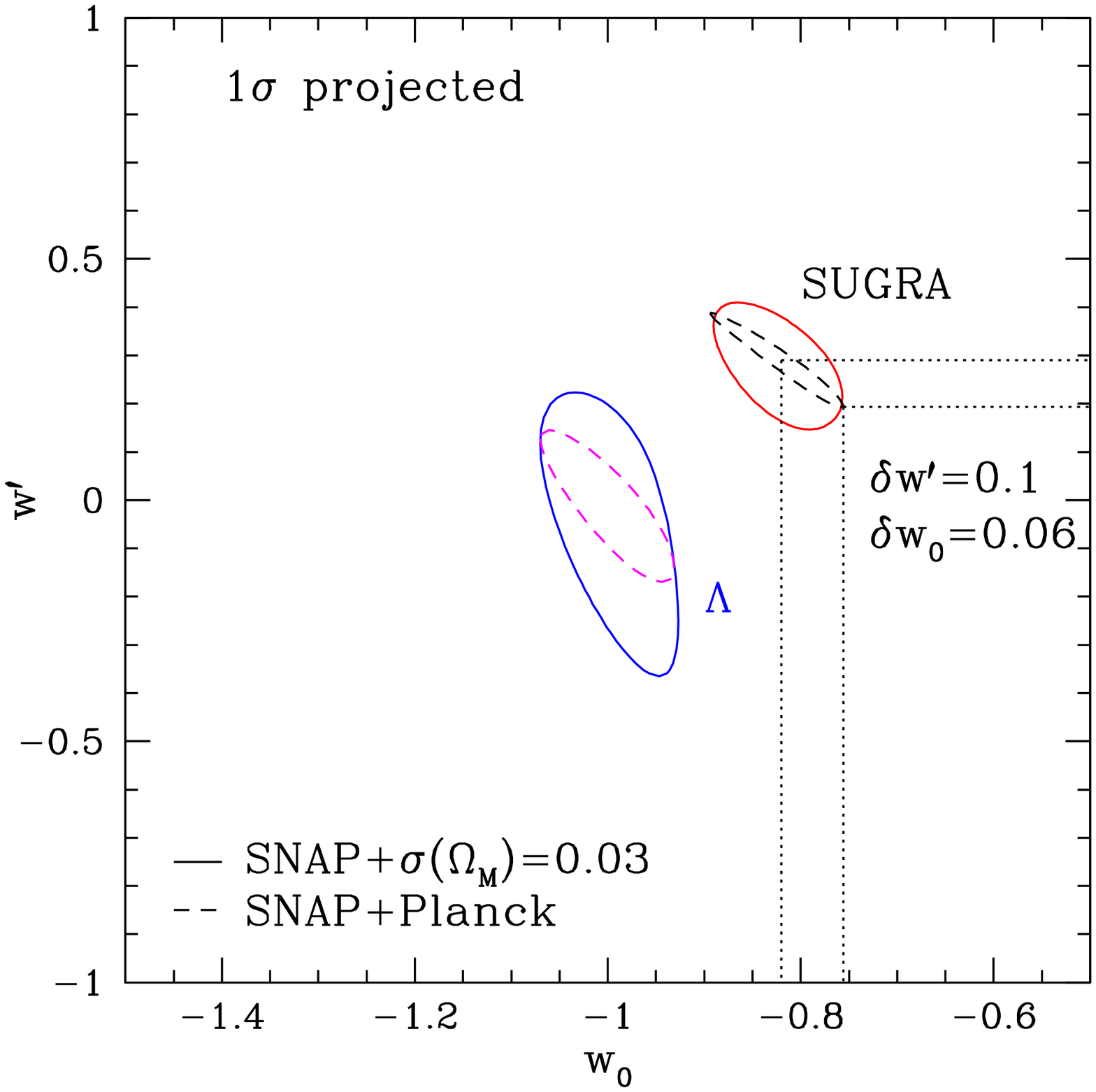}{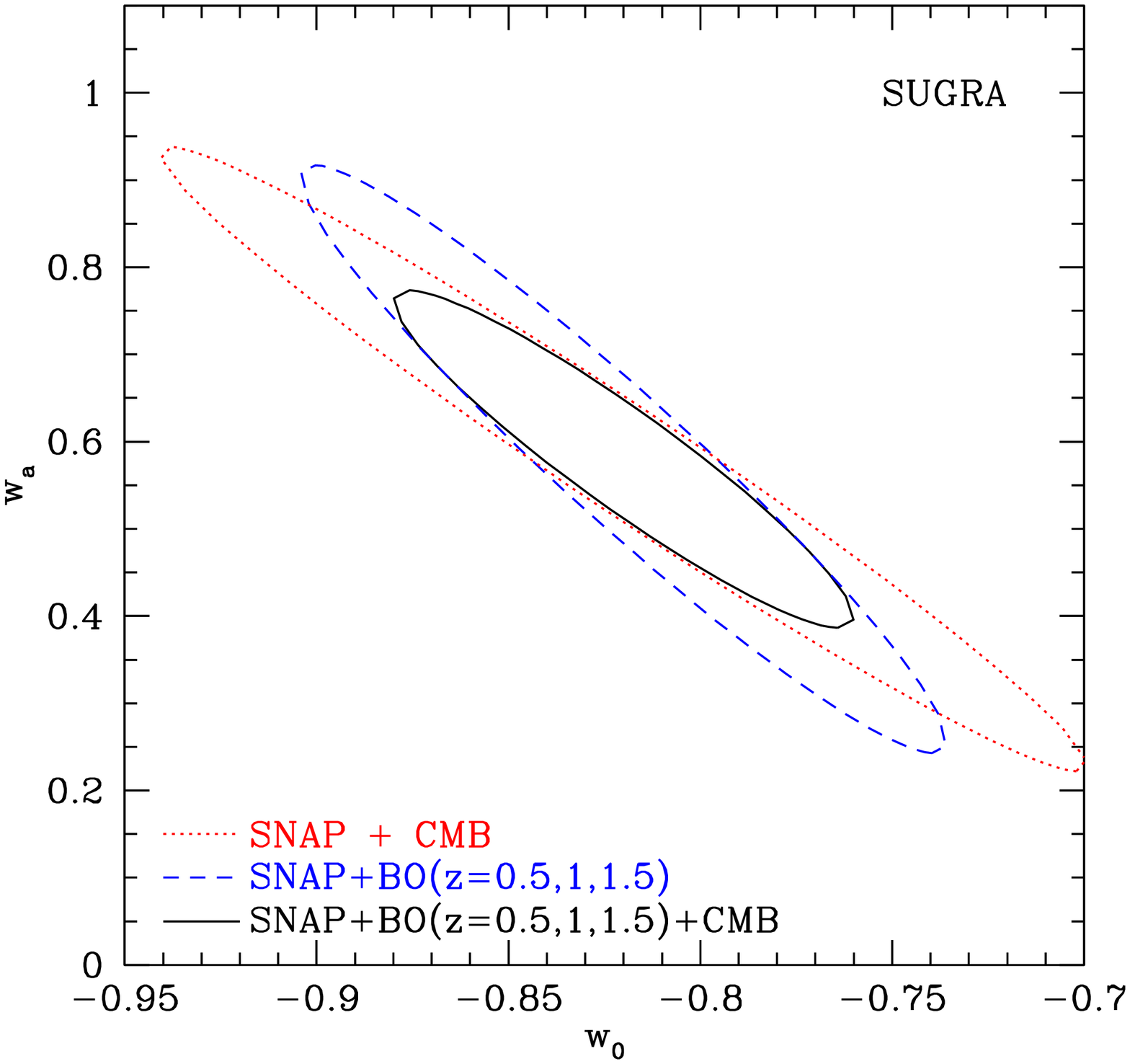}
\caption{{\it Left panel\/} -- Complementarity between SNAP supernova 
data and Planck CMB measurements greatly improves dark energy constraints 
($w'=w_a/2$) and removes the need for a prior on $\Omega_M$.  
{\it Right panel\/} -- 
Baryon oscillation measurements also provide good complementarity in 
the case of time varying equation of state. 
}
\label{fig.sncmb}
\end{figure}

Weak gravitational lensing offers another important probe.  Wide 
field, deep data such as from SNAP can constrain 
$\Omega_M$ independently and place limits on a combination of the 
present EOS $w_0$ and the time variation $w'$.  This provides a 
good crosscheck on the supernova plus CMB results, or further 
complementarity in the time varying EOS case.  See 
Linder \& Jenkins (2003) for calculations using the linear part of the 
lensing shear power spectrum and Jain \& Taylor (2003) and Bernstein 
\& Jain (2003) for shear crosscorrelation methods. 

\section{Mapping the Growth History} 

The expansion history $a(t)$ offers a clear method for mapping 
the equation of state $w(z)$, with the hope of then revealing the 
underlying physics, for example the scalar field potential $V(\phi)$. 
The quantity $w'$ is directly related to the slow roll parameter 
$V'/V$.  But as we have seen in eq.\ 3, general modifications 
of the expansion rate can lead to $w(z)$.  So we may not be able to 
uniquely interpret even very precise results.  Ideally we would like 
a second avenue to investigate dark energy, where the equation of 
state enters differently. 

Growth of structure in the universe provides such a path in theory. 
As dark energy begins to become an important fraction of the total 
energy density, it acts to shut down the growth of density perturbations 
in the matter.  Through the hierarchical process of structure formation 
this then has implications for, e.g., galaxy mass profiles and 
the abundance of galaxy clusters.  However, these objects also 
involve hydrodynamics and feedback, nonlinearities, 
and a host of sources of astrophysical confusion.  The requirement of 
accurate, well understood probes leads us to look at the largest, 
mostly linear scales. 

Data for precise measurements of effects on the matter power spectrum 
on large scales requires wide field, deep surveys.  One example is 
weak lensing surveys discussed above, which gravitationally detect even 
the dark matter.  Another involves galaxy redshift surveys.  For both, one 
seeks orders of magnitude improvement, taking for example the recent 2dF 
(two square degree field) survey and enlarging it to, say, 400dF.  
An exciting prospect 
is the KAOS (Kilo-Aperture Optical Spectrograph) instrument proposed 
for the Gemini 8-meter telescope.  This would allow simultaneous 
redshift determination of 4000 galaxies over a 1.5 square degree field 
of view.  Utilization of this facility in a Dark Energy Project 
(KAOS {\it Purple Book\/} 2003) to observe a million galaxies at 
redshifts $z\approx1$ 
and $z\approx3$ could probe dark energy by measuring the baryon 
oscillations in the matter power spectrum.  

These baryon oscillations are the analog of the acoustic peaks in the CMB 
temperature power spectrum.  They both arise from the decoupling era, 
$z\approx1100$, when density perturbations in both the baryons and 
photons could only oscillate without growing.  While the peaks and 
troughs left in the photon spectrum are large, the baryons are 
overwhelmed by cold dark matter and so only leave wiggles in the matter 
spectrum.  Since the scale of these wiggles is set by the 
physics at the decoupling era between baryons and photons, they act 
as a standard ruler to determine the ratio of the observed oscillation 
scale to the sound horizon.  This effectively measures both the 
angular distance to the redshift of the galaxies used, and the Hubble 
parameter $H(z)$ at that redshift (Blake \& Glazebrook 2003; Linder 2003b; 
Seo \& Eisenstein 2003).  Because 
the wiggles depend on well determined physics, measurements in the linear 
regime, and scales rather than amplitudes, this probe is substantially free 
from systematic uncertainties. 
The baryon oscillation method offers good crosschecks and 
complementarity with the more precise supernova and CMB methods, as 
illustrated in the right panel of Fig.\ 2. 

Another aspect of large scale structure in the linear regime is the 
growth of perturbations.  The growth factor determines the linear part 
of the matter power spectrum, and enters the nonlinear part in various 
semianalytic treatments like extended Press-Schechter formalisms or 
the halo model.  The influence of dark 
energy appears in two areas: the Hubble drag term that slows the 
linear perturbation growth and the size and evolution of the matter 
source term (see Linder \& Jenkins 2003 for more discussion).  
Figure 3 (left panel) 
shows this influence on the normalized growth $(\delta\rho/\rho)/a$.  
Note that time varying EOS models can have an 
appreciable effect at quite high redshifts.  Another point of importance 
is that dark energy models that appear degenerate with respect to the 
CMB (the constant $w_{\rm eff}$ models and the corresponding time 
varying model in brackets) can be distinguished via the growth factor. 

\begin{figure} 
\plottwo{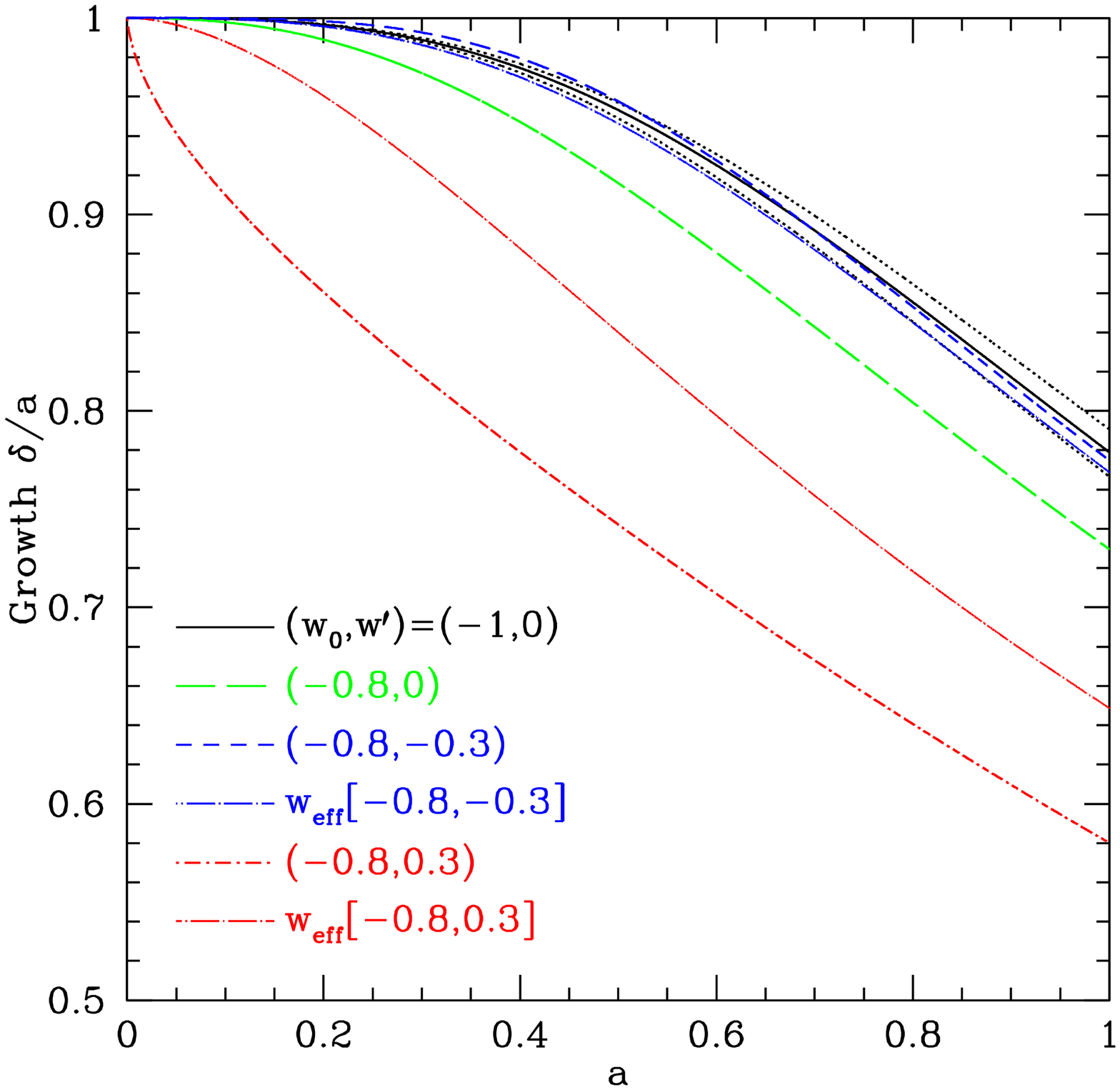}{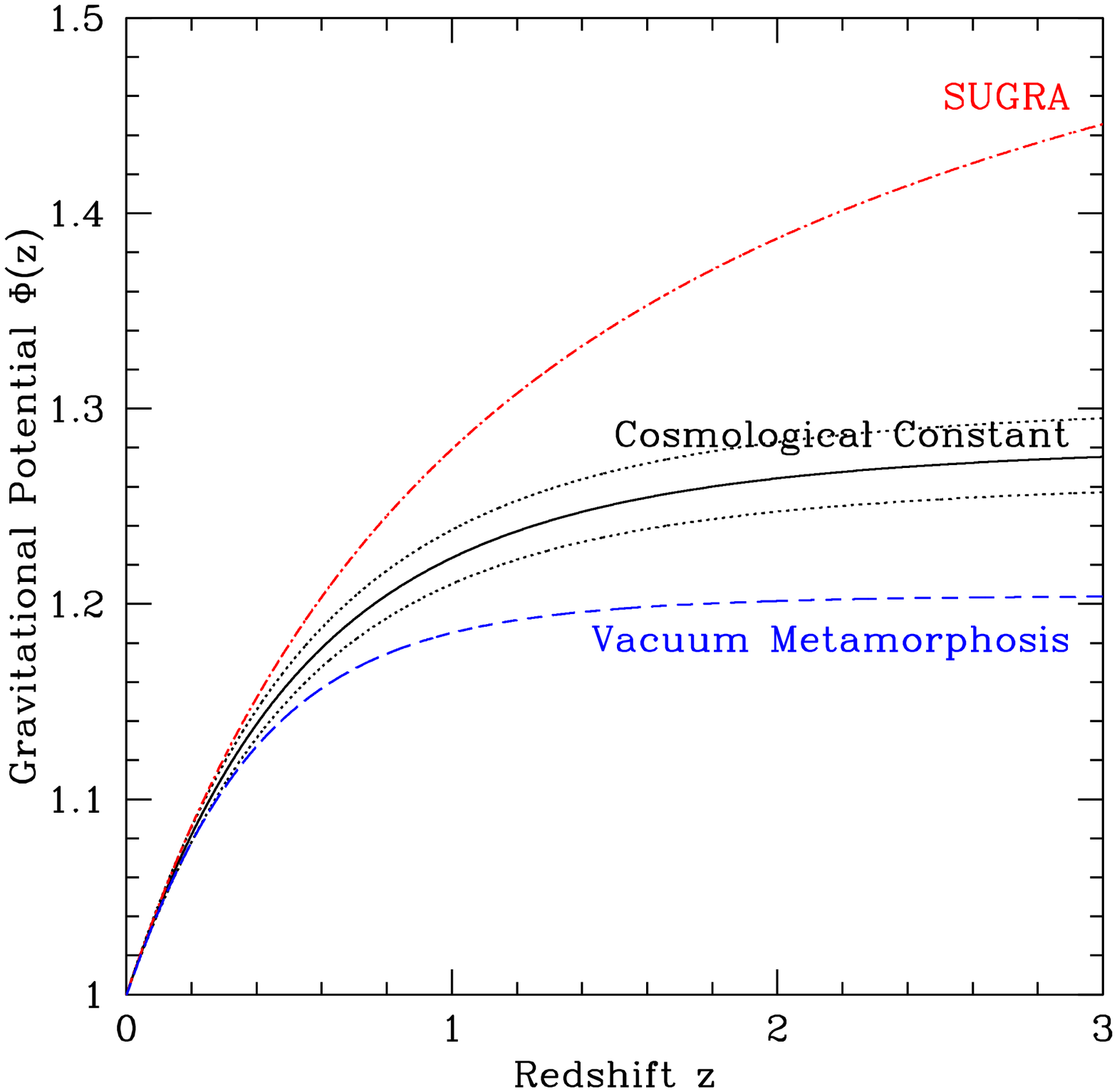}
\caption{{\it Left panel\/} -- Growth factor of linear density 
perturbations can see time variation in the dark energy equation of 
state.  Moreover, it separates the time varying models in brackets 
from the constant $w_{\rm eff}$ ones degenerate with respect to the CMB. 
{\it Right panel\/} -- Gravitational potential exhibits clear variations 
in time evolution for different dark energy models.  Dotted lines denote 
variation of $\Omega_M$ by $\pm0.02$ in the cosmological constant model. 
}
\label{fig.gro}
\end{figure}

The growth factor $\delta/a$ is directly proportional to the gravitational 
potential $\Phi(z)$, through the Poisson equation.  This not only gives 
a useful visual representation of the decay of potentials as dark energy 
begins to dominate over matter (see the right panel of Fig.\ 3), 
but is central to the integrated 
Sachs-Wolfe effect on the CMB low multipoles or large angles.  Dark energy 
parameters enter with a different dependence than for supernova distances 
or the CMB acoustic peak location.  Unfortunately the large angle region 
of the CMB suffers strongly from cosmic variance, so it is not easy to 
extract dark energy characteristics from the data, though various methods 
of crosscorrelation are proposed to try.  Nevertheless, the growth history 
in one manifestation or another offers attractive complementarity with 
the expansion history as a probe, in particular for looking at modifications 
of gravity or more complicated dark energy models that involve nonminimal 
couplings or noncanonical sound speeds. 

To attempt to use properties of galaxies or clusters as dark energy probes, 
we must understand enough to cleanly disentangle astrophysics of these 
objects from the cosmology.  This is a challenging prospect, both for theory 
and observation.  For example a shift by 10\% in the limiting mass 
threshold when counting clusters as a function of redshift is degenerate 
with a systematic bias in the EOS by 10\% (M.\ White, private 
communication).  

On the theoretical side, until recently no calibration of the cluster mass 
function (numbers of clusters as a function of mass and redshift) existed 
for models with time varying EOS.  The results of Linder \& Jenkins 2003 
for the 
highest mass clusters (least subject to nonlinearities and astrophysical 
effects) offer some hope as the mass function seems determined predominantly 
by the linear growth factor.  Indeed, it is fit to within 20\% by the 
standard Jenkins et al.\ (2001) mass formula and definite differences in the 
amount and evolution of large scale structure exist for various dark energy 
models.  But considerable research remains before we can confidently use 
galaxies and clusters directly for precision cosmology. 

\section{Conclusion} \label{sec.concl}

Dark energy poses a fundamental mystery as to what composes the majority 
of the universe, dominates its dynamics through a gravitational 
repulsion, and determines the fate of the universe.  Constraints on 
an averaged equation of state quantity from impressive efforts over 
the last five years show that it behaves roughly like a cosmological 
constant.  The precision of 15\% is already good enough that continuing 
measurements along these lines are unlikely to be able to detect a 
deviation from cosmological constant properties at 3$\sigma$.  But 
powerful next generation experiments are being designed that are 
sensitive not to a crude approximation of a constant EOS but that 
can map out the dynamical physics of a time varying equation of 
state $w(z)$.  These will also generate extraordinary astronomical 
data resources.

Mapping the expansion history to $z=1.7$ through the Type Ia supernova 
method offers great promise.  Even more valuable gains in accuracy 
and precision come from working together with crosschecking and 
complementary methods such as CMB measurements, weak gravitational 
lensing, and baryon oscillations measured in the matter power spectrum 
by large, deep galaxy surveys.  Mapping the dark energy equation of 
state will give us guideposts to high energy physics, the early 
universe, extra dimensions, or the theory of gravity, as well as 
reveal to us the true nature of the universe we live in and a picture 
of our fate. 

\acknowledgments 

I am grateful to the American Astronomical Society and National Science 
Foundation for an international travel grant.  This work was supported 
in part by the Director, Office of Science, US DOE under 
DE-AC03-76SF00098 at LBL.

\end{document}